\documentclass[9pt,twocolumn,twoside,lineno]{pnas-new}

\usepackage{hyperref}
\usepackage{pdfpages}

\newcommand{\anonymizejoby}[1]{#1}
\newcommand{\anonymizekitty}[1]{#1}
\newcommand{\anonymizelili}[1]{#1}
\newcommand{\anonymizebeta}[1]{#1}
\newcommand{\anonymizearcher}[1]{#1}

\templatetype{pnasbriefreport} 

\title{The promise of energy-efficient battery-powered urban aircraft}

\author{Shashank Sripad}
\author{Venkatasubramanian Viswanathan} 

\affil{Department of Mechanical Engineering, Carnegie Mellon University, Pittsburgh, PA 15213}

\leadauthor{Sripad} 

\authorcontributions{S.S. and V.V. conceived the idea, performed the analyses, and wrote the paper.}
\authordeclaration{No competing interests.}
\correspondingauthor{\textsuperscript{2}To whom correspondence should be addressed. E-mail: venkvis\@cmu.edu}


\begin{abstract}
Improvements in rechargeable batteries are enabling several electric urban air mobility (UAM) aircraft designs with up to 300 miles of range with payload equivalents of up to 7 passengers. We find that novel UAM aircraft consume between 130 Wh/passenger-mile up to $\sim$1,200 Wh/passenger-mile depending on the design and utilization, relative to an expected consumption of over 220 Wh/passenger-mi for terrestrial electric vehicles and 1,000 Wh/passenger-mile for combustion engine vehicles. We also find that several UAM aircraft designs are approaching technological viability with current Li-ion batteries, based on the specific power-and-energy while rechargeability and lifetime performance remain uncertain. These aspects highlight the technological readiness of a new segment of transportation. 
\end{abstract}

\dates{This manuscript was compiled on \today}
\doi{\url{www.pnas.org/cgi/doi/10.1073/pnas.XXXXXXXXXX}}

\begin{document}

\maketitle
\thispagestyle{firststyle}
\ifthenelse{\boolean{shortarticle}}{\ifthenelse{\boolean{singlecolumn}}{\abscontentformatted}{\abscontent}}{}

\dropcap{A}ircraft designed to travel up to 300 miles are currently used for various applications including mobility of passengers, cargo as well as defense and emergency services via helicopters or small planes. Recently, Urban Air Mobility (UAM) has emerged as a platform that could transform transportation in urban areas and displace activities serviced by terrestrial vehicles.  UAM concepts hinge on the development of electric vertical take-off and landing (EVTOL) aircraft. These aircraft operate using `vertiports' (similar to helipads) with no runway, making them particularly suitable for urban environments. EVTOL aircraft also present a two-to-six-fold faster means of point-to-point mobility compared to terrestrial alternatives.\cite{kasliwal_role_2019} Large investments amounting to several billion U.S. Dollars have been mobilized for this sector in 2021.\cite{Bryant_bloomberg}

Aircraft electrification enables distributed (electric) propulsion since electric motor efficiency and power density are scale-invariant, unlike combustion engines. A large number of small electric motors could be used instead of conventional combustion-based propulsion architectures with a few ($<$4) relatively large propulsion units.\cite{epstein2020aeropropulsion,stoll_drag_2014} Distributed propulsion reduces drag significantly,\cite{epstein2020aeropropulsion,stoll_drag_2014} while electric motors are about 2\textemdash3 fold more efficient than combustion engines, resulting in higher overall efficiency for electric aircraft.\cite{epstein2020aeropropulsion}

Over the last few years, several novel UAM aircraft designs have emerged, enabled by the improvements in specific energy and power associated with Li-ion batteries.\cite{epstein2020aeropropulsion}  The UAM aircraft design space is comprised of a highly diverse set of specifications for cruising distance, maximum take-off mass (MTOM), payload capacity, and rate of energy consumption. 
There are three broad categories of EVTOL aircraft: (1) Multi-rotor, similar to helicopters but with multiple rotors distributed over an aircraft, generally without a fixed-wing, (2) Lift-plus-cruise, where one set of rotors are used for take-off and landing (vertical flight) and another set for cruising, generally with a fixed-wing, (3) Vectored Thrust, generally fixed-wing aircraft where the thrust providing system of the aircraft is used both in vertical and forward flight by maneuvering the direction of thrust; can be further categorized into (3a) Tilt-rotor, where rotors used in vertical flight tilt via rotating shafts to be used in forward flight, (3b) Tilt-wing, where the tilting action is performed by wings onto which the rotors are attached, and (3c) Tilt-duct, similar to tilt-rotor but the thrust is generated by propellers that use rotors housed within cylindrical ducts, sometimes called ducted fans.

The power requirement in vertical flight is strongly influenced by the design parameter of disc loading (kg/m$\mathrm{^2}$) which is the ratio of MTOM to total rotor disc area.\cite{mcdonald2017evtol} Multi-rotors and aircraft with a larger total rotor disc area, resulting in a lower disc loading, require lower power for take-off and landing.\cite{mcdonald2017evtol} On the other hand, designs with a low total rotor disc area require high vertical flight power. Horizontal flight power requirements are influenced strongly by the lift-to-drag (L/D) ratio,\cite{mcdonald2017evtol} as shown in \textit{Materials and Methods}. Multi-rotors with large rotors cause an increase in drag leading to high power requirements during cruise. Aircraft with fixed-wings that provide lift during cruise have higher energy efficiency in horizontal flight. 

Notwithstanding the differences in power requirements due to design parameters like disc loading or L/D ratio, across all EVTOL designs, the energy consumption per unit mile traveled for take-off, landing, and hovering segments is much higher than the cruise segment. Therefore, to a first approximation, the total energy consumption per unit mile for a trip is directly proportional to the fraction of time spent in vertical flight. For fixed take-off and landing segments, once an aircraft reaches the specified flying altitude, as the cruise distance increases, the overall energy consumption per unit mile for the trip generally decreases (with an optimum at a certain cruising speed). Previous studies\cite{kasliwal_role_2019} have used fixed values for parameters such a disc loading resulting in estimates failing to describe several new aircraft designs.\cite{nathena2021architectural,bacchini2019electric}

\begin{figure}
\centering
\includegraphics[trim=0cm 0cm 1.1cm 0.9cm, clip=true, width=\linewidth]{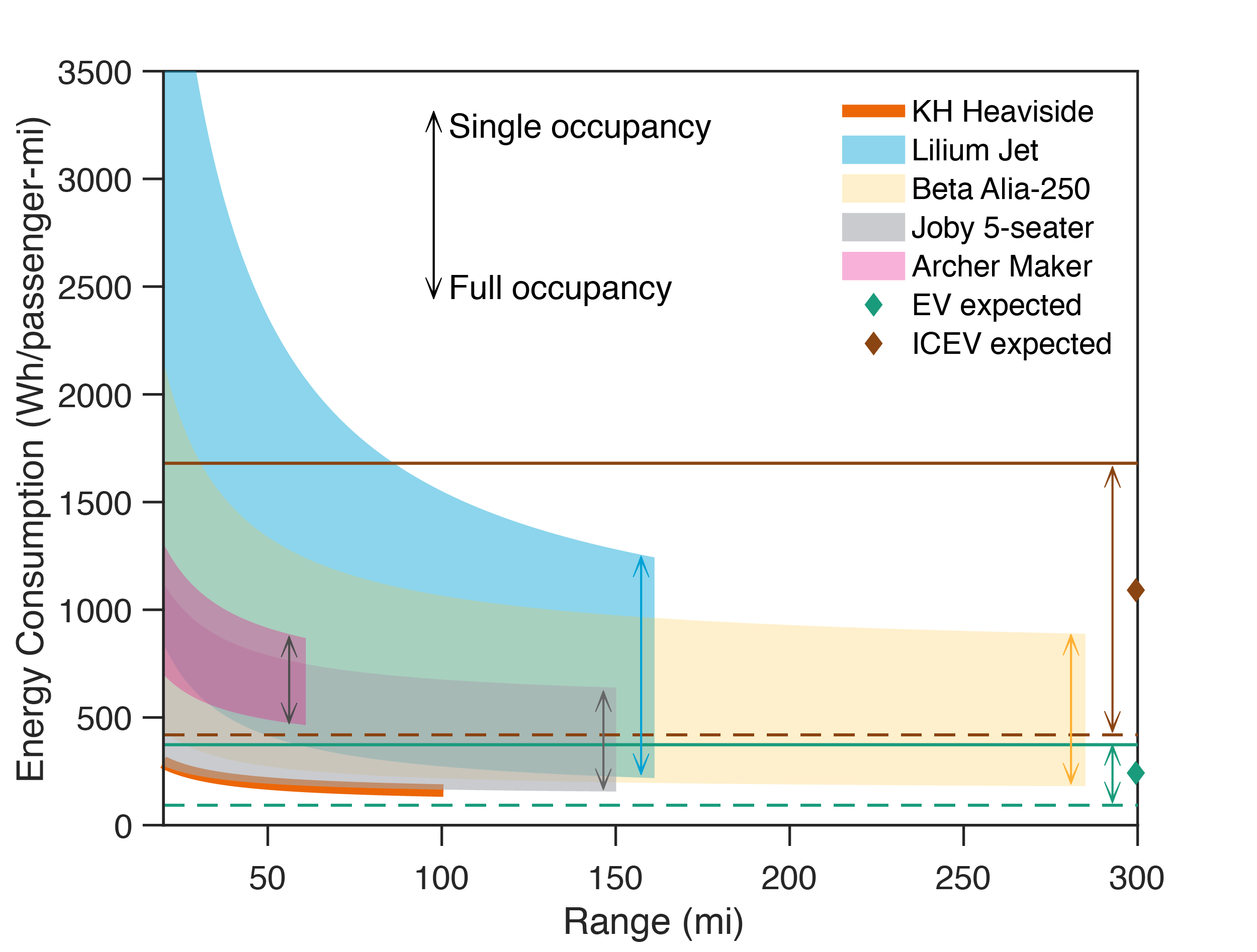}
\caption{Energy efficiency different EVTOL aircraft and terrestrial vehicles. The diamond markers represent the expected EV (Electric Vehicle) and ICEV (Internal Combustion Engine Vehicle) at an occupancy of 1.67.\cite{nhts} The energy consumption for all EVTOL aircraft is estimated at a cruising speed of 150 mi/h which is up to six fold faster than equivalent terrestrial vehicles.\cite{kasliwal_role_2019} Energy consumption for \anonymizekitty{single} passenger \anonymizekitty{KH Heaviside} is occupancy invariant. As the length of the cruise segment increases the energy consumption decreases. Fully occupied EVTOLs are equivalent or more energy-efficient than fully occupied ICEVs for flying ranges more than 70 miles or lower depending on the aircraft, while the energy consumption is similar or lower than an expected EV (223 Wh/passenger-mi) after 100 miles.}
\label{fig1}
\end{figure}

To compare the energy efficiency of terrestrial vehicles like electric vehicles (EVs) and EVTOL aircraft, certain differences between the two modes need to be accounted for. EVTOLs cover point-to-point distance without meanders, whereas, EVs travel on roads with circuitous paths resulting in a longer distance covered between the same points. Previous studies have suggested circuity factors between 1.12 and 2.10 for various countries, while the U.S. average route circuity is about 1.20.\cite{kasliwal_role_2019} Another important factor is the number of occupants or the amount of payload carried by the vehicles. In the United States, the average occupancy for light vehicles including motorcycles, cars, and light trucks has been 1.67 for over 10 years.\cite{nhts} The appropriate metric to compare energy efficiency, in this context, is the energy consumption per unit distance per unit payload carried. We use Watt-hour/passenger-mile after accounting for circuity for terrestrial vehicles. 

We choose five EVTOL aircraft respresentative of the diverse EVTOL aircraft design space, (1) \anonymizekitty{\textit{Kitty Hawk Corporation} (KH) Heaviside,} (tilt-rotor) (2) \textit{\anonymizejoby{Joby Aviation}} (\anonymizejoby{Joby}) \anonymizejoby{2021} (yet to be named aircraft), (tilt-rotor) (3) \anonymizelili{\textit{Lilium GmbH} (Lilium) Jet,} (tilt-duct), (4) \anonymizebeta{\textit{Beta Technologies} (Beta) Alia-250}, (lift-plus-cruise), and \anonymizearcher{\textit{Archer Aviation}} \anonymizearcher{(Archer) Maker}, (lift-plus-cruise/tilt-rotor) each designed to carry \anonymizekitty{one}, \anonymizejoby{five}, \anonymizelili{seven}, \anonymizebeta{six}, and \anonymizearcher{two} passengers (including the pilot, if used) while traveling \anonymizekitty{100}, \anonymizejoby{150}, \anonymizelili{172 (150 nautical miles)}, \anonymizebeta{288 (250 nautical miles)}, and \anonymizearcher{60} miles respectively. A previously developed EVTOL power consumption model is used to analyse the range, payload (passengers), and energy consumption trade-off.\cite{fredericks2018performance}  The model is described in \textit{Materials and Methods} and \textit{SI Appendix}.

\section*{Results and Discussion}
In Figure \ref{fig1}, across all aircraft considered, as the length cruise segment increases with longer flying range, the efficiency improves drastically. For a single passenger, the energy consumption of larger aircraft is generally higher, \anonymizelili{Lilium Jet}>\anonymizebeta{Beta Alia-250}>\anonymizearcher{Archer Maker}$\approx$\anonymizejoby{Joby 5-seater}>\anonymizekitty{KH Heaviside}. \anonymizelili{Lilium Jet} uses ducted fans resulting in high energy consumption for vertical flight due to the high disc loading, but as cruise length increases, the energy consumption drops rapidly compared to other aircraft due to its highly efficient cruising segment.


Figure \ref{fig1} describes the energy consumption comparison of the five aircraft with a terrestrial EV and ICEV. The EV and ICEV are assumed to have a fixed duty cycle over the travel distances analysed and are examined at single, maximum, and expected occupancy. The details of the estimates for EV and ICEV are in \textit{SI Appendix}. At median occupancy, all five aircraft are more efficient at designed range than the expected ICEV (1,000 Wh/passenger-mi). At full occupancy and designed range, all aircraft, are more efficient or equivalent to a fully occupied ICEV (420 Wh/passenger-mi). Beyond 20 mi, the \anonymizekitty{KH Heaviside} is always more efficient than an ICEV irrespective of the ICEV occupancy considered. 

On comparing the efficiency of EVs with the five aircraft, we find that the single passenger \anonymizekitty{KH Heaviside} is more efficient than an EV with one occupant at ranges greater than 20 miles and more efficient than the expected EV at ranges greater than 35 miles. Fully occupied \anonymizejoby{Joby 5-seater}, \anonymizebeta{Beta Alia-250}, and \anonymizelili{Lilium Jet} show an energy consumption of about 156, 181, and 218 Wh/passenger-mi respectively, at their designed flying range, all lower than the expected EV at 223 Wh/passenger-mi. This represents a significant energy-efficiency milestone for EVTOL aircraft highlighting the enormous efficiency gains that can be achieved via fixed-wing cruising.

\begin{figure}
\centering
\includegraphics[trim=0cm 0cm 1.1cm 0.9cm, clip=true, width=\linewidth]{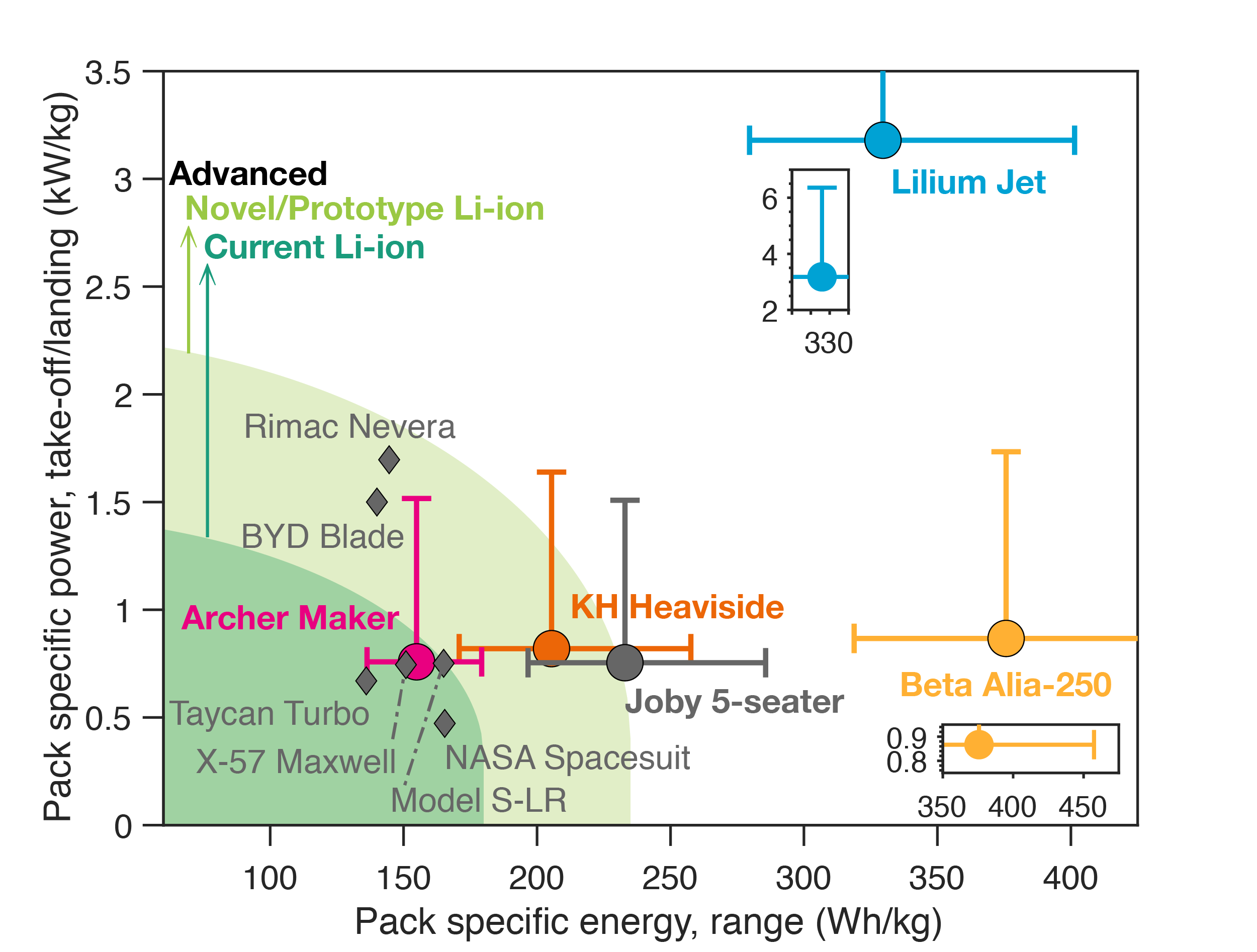}
\caption{The minimum pack specific energy and specific power (discharge) requirements for the different aircraft analyzed at an empty weight fraction (EWF) of 0.5 while the abscissa error bars indicate estimates at an EWF of 0.45 and 0.55. A cruising speed for maximum range with 30 min reserves is assumed for battery sizing. The ordinate error bars show the landing power requirement where half the battery pack has failed. Various battery packs that have been developed to date are shown and labeled as grey diamonds (dataset in \textit{SI Appendix}). `Current Li-ion' representing batteries manufactured at large scale; `Novel/prototype Li-ion' indicates chemistries and designs developed recently or for high-performance applications; `Advanced' indicates nascent pack designs that are not yet commercially available.}
\label{fig2}
\end{figure}

One of the crucial enabling factors for modern EVTOL aircraft, as noted previously, is the battery pack.\cite{epstein2020aeropropulsion} There has been tremendous progress in performance and cost of Li-ion and related battery chemistries over the last decade. However, earlier studies on EVTOL aircraft\cite{kasliwal_role_2019,bacchini2019electric} make fixed cell-level specific energy assumptions for batteries and/or no consideration of specific power thereby missing the interplay between aircraft design parameters and battery requirements. 

Given the advanced thermal management systems that exist in modern aircraft, some EVTOL manufacturers have proposed approaches to designing battery packing and management systems that are integrated with other onboard systems thereby improving the pack-level specific energy.\cite{nathena2021architectural,villanueva2021battery} The MTOM of an electric aircraft can be broadly divided into three parts, (1) payload, (2) battery weight, (3) empty weight that accounts for the weight of the aircraft structures, airframe, propulsion systems, and other on-board systems. 

In Figure \ref{fig2}, we explore the battery pack specific energy and specific (discharge) power requirements, as defined by the range and take-off/landing power demands for the five EVTOL aircraft. A comparison with several currently available battery pack designs in EVs, experimental (X) planes, and space applications is shown (see \textit{SI Appendix} for dataset). In Figure \ref{fig2}, we show three categories for battery pack technology based on technical readiness and commercial availability.

The battery pack specific energy requirements in Figure \ref{fig2} are estimated using an empty weight fraction (EWF) of 0.5, (lower than current aviation standards\cite{fredericks2018performance}) to facilitate the possible use of battery packing weight for structures and other shared functions.\cite{nathena2021architectural,villanueva2021battery} Estimates are shown using uncertainty bounds for EWF of 0.45 and 0.55 are reflected as the abscissa error bars. A lower empty weight fraction provides more weight allocation for the battery thereby reducing the required battery performance metrics.

Aircraft like the \anonymizelili{Lilium Jet} that have a high disc loading (see \textit{SI Appendix}), require higher power for take-off, landing, and hover compared to other designs. Coupled with a high MTOM, the specific power requirements for the \anonymizelili{Lilium Jet} are much higher than other aircraft, as seen in Figure \ref{fig2}. Longer flying range requires larger battery packs resulting in higher specific energy requirements for aircraft like the \anonymizebeta{Beta Alia-250}. On the other hand, low range aircraft like \anonymizearcher{Archer Maker} require much lower specific energy and such designs are feasible with current Li-ion batteries. The importance of empty weight fraction can be observed by examining the increase in specific energy required for each aircraft to accommodate a higher EWF of 0.55. The uncertainty limits for power accounts for the possibility of partial failure of the battery pack in a scenario where only 50\% of the battery pack supplies the total required power to land. The strong influence of EWF and battery pack failure on specific energy and power requirements shows that regulations could play an important role in determining the technical viability of EVTOL aircraft.

Figure \ref{fig2} emphasizes the importance of specific power being a more critical performance metric for EVTOLs which determines whether an EVTOL can safely take-off and land. On the other hand, to a first approximation, specific energy determines the operating range of the EVTOL. It should also be noted that Figure \ref{fig2} does not make provisions for degradation in performance metrics, and the values could be considered the minimum required performance at the end-of-life, especially for specific power given the ability to land is safety-critical. Other aspects related to battery behavior during high power requirements during landing is not reflected in Figure \ref{fig2}, and relevant discussions can be found elsewhere.\cite{fredericks2018performance} In the overall analysis, in Figure \ref{fig2}, we find that several EVTOL designs can achieve the promising energy efficiency shown in Figure \ref{fig1} via suitable improvements to current Li-ion batteries while charging and performance over lifetime require further investigation. This highlights the technological readiness of EVTOL aircraft, from the battery technology standpoint.

In this Brief Report, we have discussed two main details, (1) energy efficiency of EVTOL compared to terrestrial vehicles, and (2) battery requirements compared to current battery technological landscape. We noted the technological readiness in terms of battery requirements. The promise of EVTOL aircraft achieving higher energy efficiencies than equivalent terrestrial alternatives at faster travel times signals enormous implications for the emission intensity and sustainability of urban transportation.

\matmethods{The vertical flight power for open rotor aircraft is given by,
\begin{gather*}
    \mathrm{P_{vertical,open} =  \Bigg[ \dfrac{f\ W}{FoM}\sqrt{\dfrac{f\ W/A}{2\ \rho}} + \dfrac{W\ V_{climb,v}}{2} \Bigg] / \eta_{vertical}}
    \label{eqn:momVert1}
\end{gather*}
\noindent and for ducted fan aircraft is given by,
\begin{gather*}
    \mathrm{P_{vertical,ducted} =  \Bigg[ \dfrac{f\ W}{2\  FoM}\sqrt{\dfrac{f\ W/A}{\rho}} + \dfrac{W\ V_{climb,v}}{2} \Bigg] / \eta_{vertical}}
    \label{eqn:momVert2}
\end{gather*}
The figure of merit ($\mathrm{FoM}$) is the ratio between ideal and actual rotor power.\cite{Raymer} ($\mathrm{f}$) is a correction factor for interference from the fuselage, and is set to the typical value 1.03.\cite{Raymer} The disc area, A, determines the disc loading. Density of air, ($\mathrm{\rho}$) is calculated at flight altitude. ($V_{climb,v}$) is the climb rate, when held at zero corresponds to `hover' conditions. Aircraft weight ($\mathrm{W}$) is the product of MTOM and acceleration due to gravity (g). $\mathrm{\eta_{vertical}}$ is the combined efficiency of the motors and electric powertrain during vertical flight conditions.
\noindent Fixed-wing segment power\cite{Raymer} ($\mathrm{P_{fixed-wing}}$) is given by,
\begin{gather*}
    \mathrm{P_{fixed-wing} = \Big[ W\ V_v + \frac{W\ V}{[L/D]}\Big] / (\eta_{fixed-wing})}
    \label{eqn:fixedP}
\end{gather*}
$\mathrm{\eta_{fixed-wing}}$ includes the efficiency of the propellers as well. The vertical velocity component ($\mathrm{V_v}$) is zero for cruise. The lift-to-drag ratio ($\mathrm{{L}/{D}}$) and the forward velocity (V) are segment specific. The operating conditions are estimated using a minimum power condition for climb and descent. Extended Methods can be found in \textit{SI Appendix}.

\subsection*{Data Availability} Aircraft parameters and battery data is available in \textit{SI Appendix} and hosted at \href{https://github.com/BattModels/evtol}{https://github.com/BattModels/evtol}. 
}

\showmatmethods{} 

\acknow{The authors thank the EVTOL aircraft engineers, manufacturers, and industry experts who provided valuable inputs and feedback for this manuscript.}

\showacknow{} 

\bibliography{refs}

\begin{thebibliography}{10}

\bibitem{kasliwal_role_2019}
A Kasliwal, et~al., Role of flying cars in sustainable mobility.
\newblock {\em\protect\JournalTitle{Nature Communications}} \textbf{10}, 1555
  (2019).

\bibitem{Bryant_bloomberg}
{C. Bryant}, {The Skies Will Be Crowded With Flying Taxis}
  (\url{https://www.bloomberg.com/opinion/articles/2021-06-11/evtol-the-skies-will-be-crowded-with-flying-taxis-by-virgin-american-airlines})
  (2021) Accessed: 13th-June-2021.

\bibitem{epstein2020aeropropulsion}
AH Epstein, Aeropropulsion: Advances, opportunities, and challenges.
\newblock {\em\protect\JournalTitle{The Bridge}} \textbf{50} (2020).

\bibitem{stoll_drag_2014}
AM Stoll, J Bevirt, MD Moore, WJ Fredericks, NK Borer, Drag {Reduction}
  {Through} {Distributed} {Electric} {Propulsion} in {\em 14th {AIAA}
  {Aviation} {Technology}, {Integration}, and {Operations} {Conference}}.
\newblock (American Institute of Aeronautics and Astronautics, Atlanta, GA),
  (2014).

\bibitem{mcdonald2017evtol}
R McDonald, B German, evtol stored energy overview in {\em Uber Elevate
  Summit}.
\newblock (2017).

\bibitem{nathena2021architectural}
{P. Nathen, Lilium GmbH}, Architectural performance assessment of an electric
  vertical take-off and landing (e-vtol) aircraft based on a ducted vectored
  thrust concept
  (\url{https://lilium-aviation.com/files/redaktion/refresh_feb2021/investors/Lilium_7-Seater_Paper.pdf})
  (2021) Accessed: 25th-April-2021.

\bibitem{bacchini2019electric}
A Bacchini, E Cestino, Electric vtol configurations comparison.
\newblock {\em\protect\JournalTitle{Aerospace}} \textbf{6}, 26 (2019).

\bibitem{nhts}
{Federal Highway Administration}, {National Household Travel Survey}
  (\url{https://nhts.ornl.gov/}) (2017) Accessed: 25th-April-2021.

\bibitem{fredericks2018performance}
WL Fredericks, S Sripad, GC Bower, V Viswanathan, Performance metrics required
  of next-generation batteries to electrify vertical takeoff and landing (vtol)
  aircraft.
\newblock {\em\protect\JournalTitle{ACS Energy Letters}} \textbf{3}, 2989--2994
  (2018).

\bibitem{villanueva2021battery}
E Villanueva, et~al., Battery thermal management system and method (2021) US
  Patent 10,960,785.

\bibitem{Raymer}
DP Raymer, {\em Aircraft Design: A Conceptual Approach. Edition 5}.
\newblock (American Institute of Aeronautics and Astronautics), (2012).

\end{thebibliography}



\section*{Supplementary material}

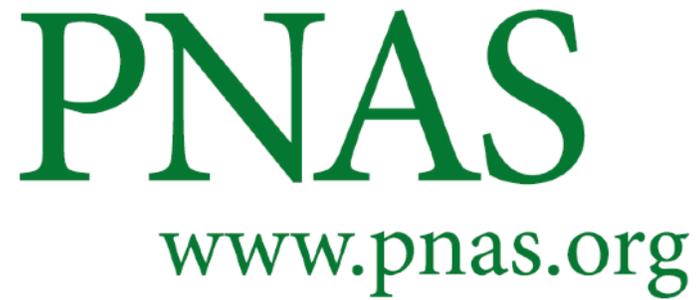

# Supplementary Information for

**The promise of energy-efficient battery-powered urban aircraft**

**Shashank Sripad, others and Venkatasubramanian Viswanathan**

**Venkatasubramanian Viswanathan.**
E-mail: venkvis@cmu.edu

**This PDF file includes:**

SI Appendix
SI References

**Supporting Information Appendix**

**Extended Methods: Power consumption model for EVTOLs**

***Minimum power condition for climb and descent.*** Forward velocity for fixed-wing climb and descent is set to the minimum power velocity ($V_{\text{MinPower}}$).(1) To do this, the equation for fixed-wing power is differentiated with respect to velocity. Setting this new expression to zero, velocity can be solved at the minimum. Some assumptions must be made about the Lift and Drag properties beyond the given L/D value, so the equation for power is expanded in the following equation,(1)

$$P_{\text{fixed-wing}} = \left[ W\ V_v + \frac{1}{2}\rho\ V^3\ S\ C_{D_0} + \frac{K\ W^2}{1/2\ \rho\ V\ S} \right] / (\eta_{\text{mech}}\ \eta_{\text{prop}})$$

Wing reference area (S) is defined by the typical wing loading (W/S). The terms $C_{D_0}$ and K are the zero-lift drag coefficient and lift-dependent correction factor respectively. They emerge from representing the total drag coefficient ($C_D$) as a function of the lift coefficient ($C_L$) as follows, (1)

$$C_D = C_{D_0} + K\ C_L^2$$

$C_{D_0}$ depends on structural aspects. Similar to the lift-to-drag ratio, K is a measure of the Drag penalty for a given Lift. Because of this, K can be represented in terms of the cruise lift-to-drag ratio, $L/D_{\text{max}}$, by combining both versions of the fixed-wing power equation, resulting in,

$$K = \frac{1}{4\ C_{D_0}\ (L/D)_{\text{max}}^2}$$

The derivative of the fixed-wing power equation with respect to forward velocity results in,

$$\frac{dP}{dV} = \frac{3}{2}\rho V^2 S\ C_{D0} - \frac{KW_{\text{TO}}^2}{1/2\rho V^2 S}$$

which can be solved for velocity as follows,

$$V_{\text{MinPower}} = \sqrt{\frac{2\ W}{\rho\ S}\sqrt{\frac{K}{3C_{D_0}}}}$$

This can be plugged back into the expanded fixed-wing power equation along with $C_{D_0}$, K, S, and $\rho$ at the appropriate flight altitude. To use the simplified equation, the lift-to-drag ratio must be adjusted for minimum power conditions. $L/D_{\text{MinPower}}$ is $0.866(L/D_{\text{max}})$, which is used for fixed-wing climb and descent. (1)

***Maximum range condition for battery sizing.*** For cruise, velocity is either set to the maximum range velocity ($V_{\text{MaxRange}}$) to perform battery sizing, or 150 mi/h for energy efficiency calculations. At different cruising velocities, the L/D is adjusted based on existing estimates relative to $L/D_{\text{max}}$.(2) Range (R), segment energy (E), average segment power (P) and velocity are related by:

$$R = \frac{E}{P}V$$

Power is constant and equal to drag times velocity at steady, level flight(1), $P = DV$.

$$R = \frac{E}{P}V = \frac{E}{D\ V}V = \frac{E}{D}$$
$$= \frac{E}{1/2\rho V^2 S(C_{D0} + KC_L^2)}$$
$$= \frac{E}{1/2\rho V^2 SC_{D0} + \frac{KW_{TO}^2}{1/2\rho V^2 S}}$$
$$= \frac{E}{1/2\rho V^2 SC_{D0} + \frac{KW_{TO}^2}{1/2\rho V^2 S}}.$$

Maximum range condition for a given energy, is given by $\frac{dR}{dV} = 0$, and rearranging to solve for velocity results in:

$$4E\rho SV(4KW_{\text{TO}}^2 - C_{D0}\rho^2 S^2 V^4) = 0$$

$$V_{\text{MaxRange}} = \sqrt{\frac{2\ W}{\rho\ S}\sqrt{\frac{K}{C_{D_0}}}}$$

**EVTOL aircraft design parameters dataset**

The design parameters used for each aircraft in the EVTOL power consumption model are discussed below:

***KH Heaviside.*** The MTOM is estimated to be 395 kg with a range of 100 miles and a single person payload of 100 kg cruising at an altitude of 1,500 m.(3) The specified cruise speed is 180 mi/h. L/D is assumed to be 18 with an estimated disc loading of 60 kg/m$^2$. The estimated energy requirement is less than 13kWh for 100 miles. The reserves are not specified.

***Joby 5-seater.*** The MTOM is 2,180 kg (4,800 lb) with a range of 150 miles and a 5 person payload of 500 kg cruising at an altitude of 450 m. The specified cruise speed is 200 mi/h. L/D is assumed to be 18 with a disc loading of 45 kg/m$^2$.(4) The estimated energy requirement is less than 150kWh for 150 miles. Additional 30 min Visual Flight Rules reserves provides a pack size requirement of 200kWh.

***Lilium Jet.*** The MTOM is 3,175 kg with a range of 150 nautical miles carrying a payload of 700 kg with a cruising altitude of 3,000 m. The specified cruise speed is about 186 mi/h (300 km/h).(5) The L/D is 18.26(5) with a disc loading of about 1,189 kg/m$^2$.(5) A form F-4 filed reveals that Lilium Jet will potentially use "battery cell chemistry based on a silicon-dominant anode combined with conventional NMC (Nickel, Manganese and Cobalt) cathodes and electrolytes" with a specific energy of "330 [Wh/kg]".(6) Since no further details are provided on the cell, the feasibility of achieving the battery pack requirements estimated in Figure 2 of the manuscript is not discussed.

***Beta Alia-250.*** The MTOM is 2,730 kg (6,000 lb) with a range of 250 miles carrying a 6 passenger load of about 600 kg with a cruising altitude of 2,500 m.(7, 8) The specified cruise speed is about 170 mi/h (274 km/h). The L/D is assumed to be 16 with an estimated disc loading of about 65 kg/m$^2$.

***Archer Maker.*** The MTOM is 1,508 kg with a range of 60 miles carrying 2 passengers. The specified cruise speed is stipulated to be 150 mi/h with an L/D of 12 and a cruising altitude of ∼609m (2,000 ft). The battery pack is specified to be 75 kWh including 15 kWh of reserves and a provision for capacity fade.(9) The specified maximum power of the battery pack is 672 kW.(9)

**Energy efficiency of terrestrial vehicles dataset**

***Internal combustion engine vehicle.*** The current U.S. average fuel economy is about 24.1 mi/gal for light-duty vehicles. Since 1 gallon of gasoline contains 33.7 kWh, the energy consumption of an ICEV is about 1,400 Wh/mi,(10) and after including the circuity of roads, it is equivalent to 1,680 Wh/mi.

***Electric vehicle.*** The U.S. department of transportation does not have current electric fleet energy consumption data, we used an average energy consumption from the U.S. Environmental Protection Agency efficiency ratings of all electric vehicles in the market.(11) We estimate an average energy consumption of about 311 Wh/mi after including 2021 model variants, in-line with other estimates(12) and about 373 Wh/mi after including circuity.

**Current battery pack performance metrics dataset**

Manuscript Figure 2 shows a number of current battery pack performance metrics which are discussed below:

***Model S-Long Range.*** The *Tesla* Model S-Long Range battery pack is estimated to have a power of 493 kW, nominal energy of 109.8 kWh with a specific energy of 165 Wh/kg.(13)

***Taycan Turbo.*** The *Porsche AG* Taycan Turbo battery pack is estimated to have a power of 460 kW, nominal energy of 93.4 kWh with a specific energy of 136 Wh/kg.(14)

***Rimac Nevera.*** The *Rimac Automobili* Nevera battery pack is estimated to have a power of 1,408 kW, nominal energy of 120 kWh weighing 830 kg.(15)

***X-57 Maxwell.*** The *NASA* X-57 Maxwell battery pack is estimated to have a power of 120 kW, nominal energy of 52.5 kWh weighing 350 kg.(16)

***BYD Blade.*** The *BYD Auto Co., Ltd.* Blade battery pack is estimated to have a specific power of 1.5 kW/kg, and a specific energy of 140 Wh/kg.(17)

***NASA Spacesuit.*** The *NASA* Spacesuit battery pack is estimated to have a power of 2.4 kW, nominal energy of 0.83 kWh weighing 5 kg.(18)

## References

1. DP Raymer, *Aircraft Design: A Conceptual Approach. Edition 5.* (American Institute of Aeronautics and Astronautics), (2012).
2. B Carson, Fuel efficiency of small aircraft. *J. Aircr.* **19**, 473–479 (1982).
3. Kitty Hawk Corporation, Heaviside (https://kittyhawk.aero/heaviside/) (2021) Accessed: 25th-April-2021.
4. A Stoll, Analysis and Full Scale Testing of the Joby S4 Propulsion System in *Transformative Vertical Flight Workshop*. (2015).
5. P. Nathen, Lilium GmbH, Architectural performance assessment of an electric vertical take-off and landing (e-vtol) aircraft based on a ducted vectored thrust concept (https://lilium-aviation.com/files/redaktion/refresh_feb2021/investors/Lilium_7-Seater_Paper.pdf) (2021) Accessed: 25th-April-2021.
6. Barry Engle, UNITED STATES SECURITIES AND EXCHANGE COMMISSION Washington, D.C. 20549 FORM F-4 REGISTRATION STATEMENT UNDER Lilium B.V. (https://www.sec.gov/Archives/edgar/data/0001855756/000110465921061848/tm2111158-12_f4.htm#tBOLA) (2021) Accessed: 12th-May-2021.
7. Beta Technologies, Alia-250 Aircraft (https://www.beta.team/aircraft/) (2021) Accessed: 25th-April-2021.
8. E. Adams, WIRED, A New Air Taxi Model Takes Design Cues From a Far-Flying Bird (https://www.wired.com/story/new-air-taxi-design-cues-far-flying-bird/) (2020) Accessed: 25th-April-2021.
9. Archer Aviation, Maker (https://www.archer.com/maker) (2021) Accessed: 25th-April-2021.
10. U.S. Department of Transportation, Bureau of Transportation Statistic, Average Fuel Efficiency of U.S. Light Duty Vehicles (https://www.bts.gov/content/average-fuel-efficiency-us-light-duty-vehicles) (2019) Accessed: 25th-April-2021.
11. M. Kane, InsideEVs, Electric Car Energy Consumption (EPA) Compared (https://insideevs.com/reviews/343702/electric-car-energy-consumption-epa-compared-april-1-2019/) (2019) Accessed: 25th-April-2021.
12. A Kasliwal, et al., Role of flying cars in sustainable mobility. *Nat. Commun.* **10**, 1555 (2019).
13. U.S. Environmental Protection Agency, Certificate Summary Information for TESLA, INC. 2021 model year test group MTSLV00.0L2S evaporative family not listed (https://iaspub.epa.gov/otaqpub/display_file.jsp?docid=51336&flag=1) (2020) Accessed: 25th-April-2021.
14. U.S. Environmental Protection Agency, Certificate Summary Information for PORSCHE AG 2021 model year test group MPRXV00.0EVT evaporative family not listed (https://iaspub.epa.gov/otaqpub/display_file.jsp?docid=51742&flag=1) (2020) Accessed: 25th-April-2021.
15. Rimac Automobili, Rimac Automobili, Nevera (https://www.rimac-automobili.com/nevera/) (2020) Accessed: 3rd-June-2021.
16. JC Chin, SL Schnulo, TB Miller, K Prokopius, J Gray, Battery performance modeling on maxwell x-57 in *AIAA Scitech 2019 Forum*. (AIAA Aviation, American Institute of Aeronautics and Astronautics), (2019).
17. XG Yang, T Liu, CY Wang, Thermally modulated lithium iron phosphate batteries for mass-market electric vehicles. *Nat. Energy* **6**, 176–185 (2021).
18. E Darcy, Passively thermal runaway propagation resistant battery module that achieves> 190 wh/kg (https://ntrs.nasa.gov/api/citations/20160003490/downloads/20160003490.pdf) (2016).



\end{document}